\documentclass[nofootinbib, aps,twocolumn,showpacs,preprintnumbers,amssymb,superscriptaddress,10pt]{revtex4-1}
\pdfoutput=1

\usepackage{tikz}
\usepackage{tikz-3dplot}
\usetikzlibrary{arrows}
\usetikzlibrary{hobby}
\usepackage{caption}

\usepackage{amsmath,amssymb,amsfonts,graphicx,epsfig,todonotes}
\usepackage[noconfig]{refstyle}
\usepackage[hyperfootnotes=true, linktocpage=true, citecolor=blue, linkcolor=blue, urlcolor=Maroon, filecolor=Maroon]{hyperref}

\usepackage{multirow}
\usepackage{adjustbox}
\usepackage{subcaption}
\usepackage{graphics}
\usepackage{tikz}
\usepackage{tikz-3dplot}
\usetikzlibrary{shapes.geometric,arrows}
\usetikzlibrary{hobby}
\usepackage{caption}

\tikzstyle{layerpink} = [rectangle, rounded corners, minimum width=0.2cm, minimum height=1cm,text centered, draw=black, fill=red!30]

\tikzstyle{layerblue} = [rectangle, rounded corners, minimum width=0.2cm, minimum height=1cm,text centered, draw=black, fill=blue!30]

\tikzstyle{layergreen} = [rectangle, rounded corners, minimum width=0.2cm, minimum height=1cm,text centered, draw=black, fill=green!30]

\tikzstyle{arrow} = [thick,->,>=stealth]

\let\eqref=\relax
\newref{eq}{name={},Name={Eq.~},names={eqs.~},Names={Eqs.~},rngtxt={-},refcmd=(\ref{#1})}
\newref{tab}{name={Table~},Name={Table~},names={tables~},Names={Tables~}}
\newref{sec}{name={Section~},Name={Section~},names={sections~},Names={Sections~}}
\newref{fig}{name={Figure~},Name={Figure~},names={figures~},Names={Figures~}}
\numberwithin{equation}{section}

\newcommand{\comment}[1]{}

\newcommand{\IP}{\mathbb{P}}

\newcommand{\IR}{\mathbb{R}}

\newcommand{\IZ}{\mathbb{Z}}

\begin{document}
\title{The World in a Grain of Sand: Condensing the String Vacuum Degeneracy}
\author{Yang-Hui He}
\affiliation{London Institute, Royal Institution of GB, 21 Albemarle St., London W1S 4BS
}
\affiliation{Merton College, University of Oxford, OX1 4JD, UK}
\affiliation{Department of Mathematics, City, University of London, London EC1V0HB, UK}
\affiliation{School of Physics, NanKai University, Tianjin, 300071, P.R. China}
\email[]{hey@maths.ox.ac.uk}

\author{Shailesh Lal}
\affiliation{Faculdade de Ciencias, Universidade do Porto, 687 Rua do Campo Alegre, Porto, Portugal}
\email[]{shailesh.hri@gmail.com}

\author{M.~Zaid Zaz}
\affiliation{Department of Physics \& Astronomy, University of Nebraska, Lincoln, NE 68588-0299, USA}
\email[]{zzaz2@huskers.unl.edu}

\preprint{LIMS-2021-12}

\begin{abstract}
We propose a novel approach toward the vacuum degeneracy problem of the string landscape, by finding an
efficient measure of similarity amongst compactification scenarios.
Using a class of some one million Calabi-Yau manifolds as concrete examples, the paradigm of few-shot
machine-learning and Siamese Neural Networks represents them as points 
in $\mathbb{R}^3$ where the similarity score between two manifolds is the Euclidean distance between 
their $\mathbb{R}^3$ representatives.
Using these methods, 
we can compress the search space for exceedingly rare manifolds to within one
percent of the original data by training on only a few hundred data points. We also demonstrate how
these methods may be applied to characterize `typicality' for vacuum representatives.
\end{abstract}


\maketitle

\section{Introduction and Summary}
The biggest theoretical challenge to string/M-theory being ``the theory of everything'' is the proliferation of possible low-energy, 4-dimensional solutions akin to our universe.
The plethora of possibilities in reducing the high space-time dimensions - where gravity and quantum field theory unify - gives rise to such astronomical numbers as the often-quoted $10^{500}$ in compactification scenarios \cite{Kachru:2003aw} as well as more recent and much larger estimates \cite{Taylor:2015xtz,Halverson:2017ffz}.
While constraints such as {\it exact} Standard Model particle spectrum place severe reduction on the allowed landscape of vacua 
\cite{Braun:2005nv,Bouchard:2005ag,Gmeiner:2005vz,Candelas:2007ac,Anderson:2011ns,Cvetic:2020fkd}, such reductions are typically on the order of 1 in $10^{10}$ and are but a drop in the ocean.

Confronted with this vastness, a key resolution would be the identification of a {\it measure} on the landscape, so that oases of phenomenologically viable universes are favoured, while deserts of inconsistent realities are slighted.
Such statistical approaches were undertaken in \cite{Douglas:2003um}, even before exact string standard models were found.
Nevertheless, finding such a measure, giving how {\it close} one compactification scenario is to another - whereby giving hope of vastly {\it reducing} the degeneracy of the string landscape - still remains a conceptual and computational puzzle.

The zeitgeist of artificial intelligence (AI) has engendered the almost inevitable 
introduction of machine learning (ML) in the exploration of the landscape of vacua
\footnote{Indeed, the current state of our knowledge of the string landscape 
is very similar to 
a typical semi-supervised learning problem, where one has at hand a minimal 
amount of human 
expert labeled data
and the bulk of the data is unlabeled. 
The methods outlined in this paper, especially in
Section \ref{sec:typicalcicy3}, may also be regarded as a natural
starting point for such analyses.}
\cite{He:2017aed,He:2017set,Carifio:2017bov}
(q.v.~\cite{Krefl:2017yox,Ruehle:2017mzq,Parr:2019bta,Otsuka:2020nsk}), and indeed, of 
mathematical structures 
in general \cite{He:2021oav} (q.v.~\cite{Ruehle:2020jrk,He:2020mgx} for reviews).
The typical approach adopted in these studies is to feed landscape data into an ML framework and pose questions either
as classifications (e.g., \cite{He:2017aed,Erbin:2020srm,Erbin:2020tks,Parr:2019bta,Deen:2020dlf}) or regressions (e.g., \cite{Ashmore:2019wzb}), 
or to use reinforcement learning \cite{Halverson:2019tkf,Harvey:2021oue,Constantin:2021for,Abel:2021rrj} 
to formulate optimal 
strategies for arriving at both top down and bottom up models of particle phenomenology.

This letter shows, using concrete examples, that ML provides a {\it natural and direct incursion onto the degeneracy problem}, using the powerful paradigm of {\it similarity}, via so-called Siamese neural networks (SNNs) \cite{bromley1993signature,hadsell2006dimensionality}, an architecture precisely designed for similarity of elements of a dataset 
 \footnote{Recently \cite{PhysRevResearch.2.033499}, SNNs were used to characterize symmetry in physical systems.}. 
In addition, SNNs possess two powerful
properties of great help in our analyses of landscape exploration: (1) \textit{few shot learning}, which evades the need for significant amount of training data, 
and (2)
\textit{supervised clustering} 
\cite{Eick2004}, which explicitly 
realizes the similarity measure.
In particular, the 
the NN learns a representation of input data into an embedding space where ``similar'' points are close together with respect to the usual Euclidean distance
\footnote{The clustering is `supervised' because
as instances of similar and dissimilar pairs are 
explicitly supplied to the ML algorithm. This is in contrast, for instance, 
to the unsupervised clustering of \cite{Otsuka:2020nsk}.}
.

This {\it similarity principle} is our saving grace from the lack of 
a {\it vacuum selection principle}: given two compactification scenarios, a similarity score gives a {\it measure} on the landscape.
Thus, whether a vacuum solution is phenomenologically viable can be decided by fiat and the ``des res'' \cite{Candelas:2007ac} of our universe can be selected, whereby compressing the vastness of the landscape into {\it typical representatives}.

\begin{figure*}[t!!]
    \includegraphics[width=\textwidth]{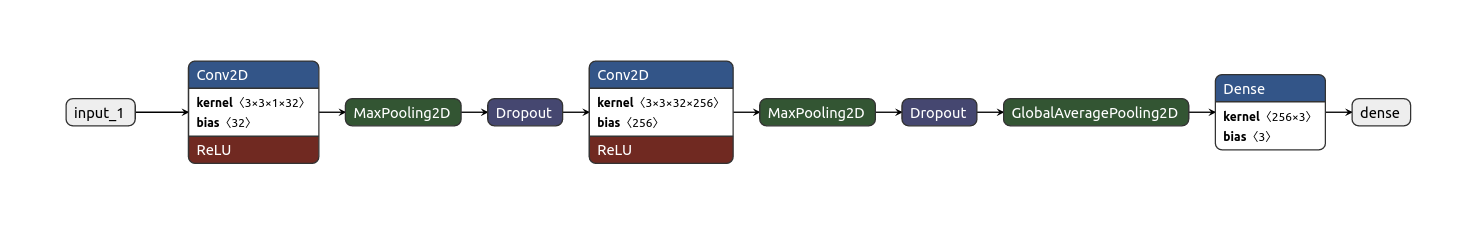}
    \caption{The `features' network. The convolutional layers enable the extraction of local features
    from CICY images.}
    \label{fig:features}
\end{figure*}

We provide here
a concrete proof-of-concept of this idea using explicit data, viz., what was referred to as ``landscape data'' in \cite{He:2017aed,He:2017set} We will focus on the so-called complete intersection Calabi-Yau (CICY) 
manifolds 
in complex dimension three
\cite{CANDELAS1988493,Green_1989,Gagnon1994ANEL,Hubsch:1992nu} and
four \cite{Gray:2013mja,Gray:2014fla}, which 
are chosen for their deep relevance 
to foundational problems in algebraic geometry and string theory
\cite{Anderson:2007nc,Anderson:2012yf,Anderson:2013xka,Deen:2020dlf} as well as the proven effectiveness
of AI/ML methods in analyzing these datasets
\cite{He:2017aed,He:2017set,He:2020lbz,Bull:2018uow,Bull:2019cij,Erbin:2020srm,Erbin:2020tks}.
These two datasets will be our representative ``landscape"
\footnote{Indeed, apart from their direct relevance to string phenomenology -- the Hodge numbers determine
the spectrum of massless fermions in the string compactification -- these datasets also explicitly realize
the general spirit of the landscape problem. The computation of the Hodge numbers is a complicated problem
whose difficulty is further exacerbated by the sheer number of cases for which this must be done. This, in
a nutshell, is the string landscape problem.}.
By clustering with similarity, one may hope to identify subsets
of data which are very likely to yield a given topology, and conversely exclude those subsets which are not.
In short, we are attempting to {\it few-shot learn the string landscape}.

This letter is organized as follows.
We begin by introducing the landscape data and their representations in \S{II} and how SNNs address them in \S{III}. The results are presented in \S{IV} and conclusions with outlook, in \S{V}.


\section{Representing the Landscape}
\label{sec:cicydataset}
The key point to our methodology 
is that a CICY can be realized as an integer matrix, which encodes the (multi-)degrees of the defining polynomials in some appropriate ambient space, see e.g. the
recent textbook \cite{he2021calabi}. While we relegate details to 
Appendix \ref{ap:histo},
the upshot is that for the purposes of computing topological quantities, a CICY is the matrix
\begin{equation}\label{eq:cicyconf}
\begin{array}{ccc}
    M = 
      \left[\begin{array}{cccc}
      q_{1}^{1} & q_{1}^{2} & \ldots & q_{1}^{k} \\
      q_{2}^{1} & q_{2}^{2} & \ldots & q_{2}^{k} \\
      \vdots & \vdots & \ddots & \vdots \\
      q_{m}^{1} & q_{m}^{2} & \ldots & q_{m}^{k} \\
      \end{array}\right]\,
      &
      &
      \begin{array}{l}
      \sum\limits_{i=1}^m  \sum\limits_{j=1}^k q_{i}^{j} = k+D+m \ ,
      \\
      \end{array}
      \end{array}
\end{equation}
where $q_i^j \in \mathbb{Z}_{\geq 0}$ and 
$D$ is the (complex) dimension of the manifold $M$, which for us will be 3 or 4, to whcih  we will refer as CICY3 and CICY4 respectively.

One of the key problems in algebraic geometry, and in parallel, in string theory, is to compute {\it topological quantities} \footnote{
Another reason why so much work so far has been focused on topological quantities is that no explicit compact Calabi-Yau metric has been found so far. We refer the reader to the recent interesting numerical
work \cite{Larfors:2021pbb} in this regard.
} from $M$.
Such topological quantities will govern such important properties such as fundamental standard-model particles. Indeed, the paradigm of string theory is that the geometry of compactification manifolds such as $M$ determines the physics of the macroscopic $(3+1)$-dimensions of spacetime.
The most famous of such a topological quantity is the so-called Hodge number, a complex generalization of Betti numbers which count the number of ``holes'' of various dimension in $M$.
There are multiple Hodge numbers for Calabi-Yau manifolds of dimension $D$
and since the early days of string phenomenology, these quantities have been interpreted as dictating the particle content of the compactification of string theory to low-energy standard model \cite{Candelas:1985en}.
We will largely focus on the positive integer $h^{1,1}$ throughout this letter.
Thus, the model for our ``landscape'' will be labelled data of the form
\begin{equation}\label{eq:CICYlandscape}
    (q^i_j) \longrightarrow h^{1,1} \ ,
\end{equation}
where the criterion for similarity $\sim$ is 
\begin{equation}
    q^{(A)} \sim q^{(B)} \quad\text{iff}\quad h^{1,1}_{(A)} = h^{1,1}_{(B)}\, \ ,
\end{equation}
between two CICY matrices labeled by $(A)$ and $(B)$.
We emphasize that our methodology is general and one could choose any other quantity of geometrical 
and phenomenological interest to label the data.

The demographics of the CICY datasets are outlined in the Supplementary Materials (Section 
and Figs there). We note here that 
the CICY3 dataset consists of 7890 matrices corresponding to 18 distinct values of $h$
while the CICY four-fold dataset is much larger, with 
905684 non-trivial entries and 23 distinct values of $h$.
An important hindrance in the study of either dataset is its extreme skewness, with 
the tails of allowed $h$ values sparsely populated, while the middle is densely populated. Indeed, this
skewness characterizes every known landscape dataset
\cite{He:2020mgx}. 

A possible way of addressing this skewness is to construct synthetic data for 
the sparely populated classes \textit{a la } \cite{Bull:2018uow}. 
Few shot learning enables us to go in a complementary direction, where we 
aggressively reduce the number of elements we need for training, even for densely populated classes.
Indeed, to learn the 7890 elements of CICY3 and 905684 elements of CICY4 we 
draw on merely 2.67\% and 0.62\% of the full datasets, respectively
\footnote{These amounts are split into Train and Validation subsets for the SNN training. More details
are in the Appendix \ref{app:traintest}.}. 

Finally, since the CICY matrices have variable shape, with entries ranging from $1\times 1$ (the famous quintic three-fold) to $15\times 18$, we
uniformize the input data by resizing each
matrix to a uniform size $n\times n$ as described in Appendix \ref{app:feature_engg}.
This uniformization further removes explicit information about the number of rows and columns
in a given CICY matrix. Since a large portion of the datasets are \textit{favorable}, i.e. the
Hodge number $h^{1,1}$ equals the number of rows in the matrix, this step also guards against
the SNN learning spurious correlations between the matrix size and Hodge number. As a further
check, we find similar results when uniformizing bi-linear interpolation
\cite{press1988numerical} which is a completely different approach from padding and washes out
any favourability information.

\section{Methodology}
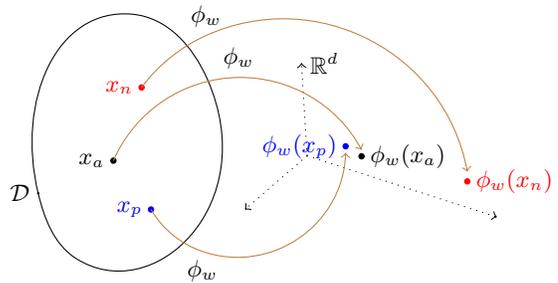
\begin{figure}
\centering
\tdplotsetmaincoords{60}{120}
\begin{tikzpicture}[scale=0.50,tdplot_main_coords,axis/.style={->},thin, use Hobby shortcut]
\draw[axis, dotted] (-0.3, 0, 0) -- (3, 0, 0) node [right]{};
\draw[axis,dotted] (-0.3, 0, 0) -- (0, 6, 0) node [right]{};
\draw[axis,dotted] (-0.3, 0, 0) -- (0, 0, 3) node [right] {$\mathbb{R}^d$};
\filldraw[black] (10,0,5) circle (2pt) node[anchor=east]{$x_a$};
\filldraw[blue] (8,0,2.5) circle (2pt) node[anchor=east]{$x_p$};
\filldraw[red] (8.5,0,6.5) circle (2pt) node[anchor=east]{$x_n$};

\filldraw[black] (2,3,2) circle (2pt) node[anchor=west]{$\phi_w(x_a)$};
\filldraw[blue] (2.5,2.8,2.5) circle (2pt) node[anchor=east]{$\phi_w(x_p)$};
\filldraw[red] (8.5,10,6.5) circle (2pt) node[anchor=west]{$\phi_w(x_n)$};
 \path
  (14,0,6) coordinate (z0)
  (12,0,3) coordinate (z1)
  (5,0,5) coordinate (z2)
  (9,0,9) coordinate (z3);
  \draw[closed] (z0) .. (z1) .. (z2)..(z3);
  
 \filldraw[black] (14,0,6) circle (0.5pt) node[anchor=east]{$\mathcal{D}$};
 
  \path
  (8,0,2.5) coordinate (p)
  (7,1,1) coordinate (inter)
  (2.5,2.8,2.3) coordinate (phip);
  \draw[style={->},brown] (p) ..(inter).. (phip);
  \filldraw[black] (7,1,1) circle (0.0pt) node[anchor=north]{$\phi_w$};
  
\path
 (10,0,5) coordinate (p)
 (12,5,10) coordinate (inter)
 (2,3,2.2) coordinate (phip);
 \draw[style={->},brown] (p) ..(inter).. (phip);
\filldraw[black] (12,5,10) circle (0.0pt) node[anchor=south]{$\phi_w$};

\path
 (8.5,0,6.5) coordinate (p)
 (12,4,11) coordinate (inter)
 (8.5,10,6.8) coordinate (phip);
 \draw[style={->},brown] (p) ..(inter).. (phip);
\filldraw[black] (12,4,11) circle (0.0pt) node[anchor=south]{$\phi_w$};
\end{tikzpicture}
\caption{The mapping $\phi_w:\mathcal{D}\rightarrow\mathbb{R}^d$ 
learnt by the features network. Similar images $A,\,P$ 
map close together while the dissimilar image $N$ maps far away.}
\label{siamese}
\end{figure}

As mentioned in the Introduction, there is tremendous difficulty but uttermost importance in defining an appropriate distance in the landscape, even theoretically, so as to identify the ``typical'' vacuum or the ``similarity'' between vacua. The key element with which an SNN solves this problem is
is a so-called {\it Features Network} (FN). This implements a map $\phi_w$
from elements of a 
dataset $\mathcal{D}$ to $\mathbb{R}^d$. Here $w$ denote the weights and biases of the FN and we set $d=3$ below. The desired property of the map $\phi_w$, visualized in Figure \ref{siamese} 
is that similar elements of $\mathcal{D}$ are mapped close together, and dissimilar elements 
are mapped far apart.
The $w$ are determined by extremizing a loss function dependent on 
\begin{equation}\label{eq:euclideandistance}
    d_w\left(x_1,x_2\right) \equiv \left(\phi_w\left(x_1\right)-\phi_w\left(x_2\right)\right)^2\,,
\end{equation}
the squared Euclidean distance between the representative points of data elements $x_1$ and $x_2$.
There are multiple options for the loss function, starting with the original 
approach of \cite{bromley1993signature,hadsell2006dimensionality}, and we adopt the 
\textit{triplet loss} function given by \cite{chechik2010large,schroff2015facenet}
\begin{equation}\label{eq:contrastiveloss}
    \mathcal{L}\left(w\right) = \max\left\lbrace d_w(x_a,x_p)-d_w(x_a,x_n)+1,0\right\rbrace\,,
\end{equation}
where $x_a$ is a reference `anchor' CICY matrix, $x_p$ is a `positive' CICY matrix with the same
$h$ as $x_a$, and $x_n$ is a 
`negative' CICY matrix with a different $h$. 
Minimizing this loss in $w$ leads to learning an embedding $\phi_w$
such that similar data cluster together and the dissimilar data are pushed apart. 
The $d_w$ in \eqref{euclideandistance} 
is then interpreted as our desired similarity score.

The input data of $n\times n$ real values matrices 
have 
a natural interpretation as pixelated images, where the $(i,j)$-th pixel is coloured according to $q^i_j$ in grey-scale.
Our features network, Figure \ref{fig:features},
exploits this ``image'' representation
by incorporating elements
from computer vision architectures, principally, convolutional layers. The resulting network is called a 
convolutional neural network or a ConvNet, and is briefly reviewed in the Appendix \ref{app:convnet}. 
\paragraph*{Note:} We may also, inspired
by \cite{Erbin:2020srm,Erbin:2020tks},
design the features network using simplified versions of the Inception
\cite{szegedy2015going,szegedy2017inception} and Residual \cite{he2016deep} blocks.
This yields comparable results, and we focus on the ConvNet for
simplicity.
\section{Results \& Discussion}
\begin{figure}[]
    \centering
    \includegraphics[width=0.3\textwidth]{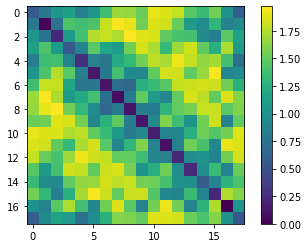}
    \caption{The similarity score for three-folds.}
    \label{fig:score_h11}
\end{figure}

\begin{figure}[]
    \centering
    \includegraphics[width=0.3\textwidth]{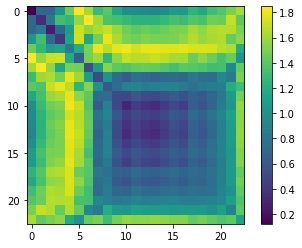}
    \caption{The similarity score for four-folds.}
    \label{fig:score_h11_4fold}
\end{figure}

We now evaluate the trained SNN by computing similarity scores across the test set, which, we recall 
are
$97.82\%$ and $99.38\%$ of the entire CICY3 and CICY4 data respectively.
Mean similarity scores 
for each pair of Hodge numbers $h^{1,1}$ from these test sets are displayed in 
Figures \ref{fig:score_h11} and \ref{fig:score_h11_4fold}.
We see that the similarity scores along the diagonal (i.e. for matrices belonging to
the same $h^{1,1}$) are concentrated close to 0, while scores for dissimilar matrices are 
concentrated away from 0, which was indeed our criterion for the putative similarity
measure.
Put together, these figures explicitly demonstrate \textit{few shot learning}; 
the SNNs have been trained on extremely sparse data, sometimes just 3 from each
class (see Tables \ref{tab:cicy3exclusion}
and \ref{tab:cicy4exclusion}). Despite this extreme paucity of training data, 
the SNNs learn a similarity score that is representative of the full datasets.
\begin{figure}[]
    \centering
    \includegraphics[width=0.45\textwidth]{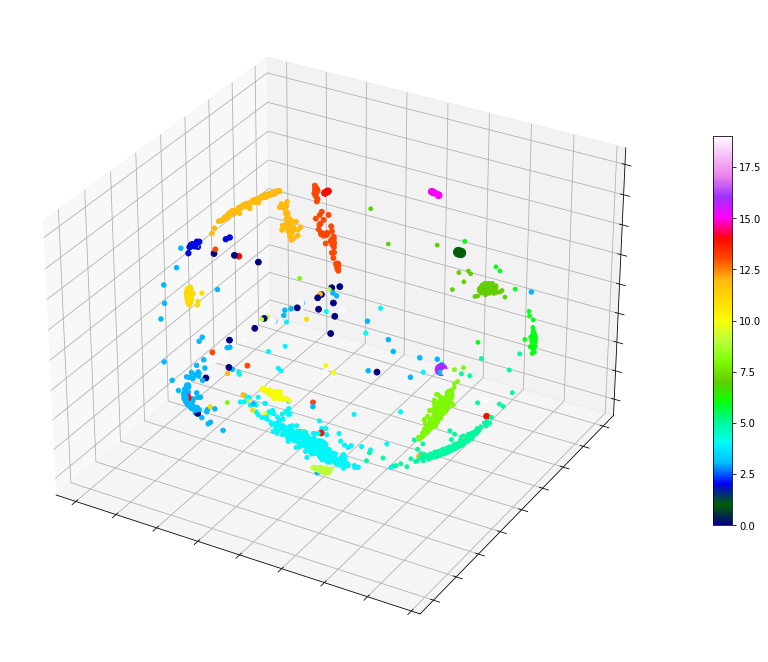}
    \caption{CICY Three-folds visualized by the SNN. The color scheme is $h^{1,1}$.}
    \label{fig:cicy_cluster}
\end{figure}
\begin{figure}[]
    \centering
    \includegraphics[width=0.45\textwidth]{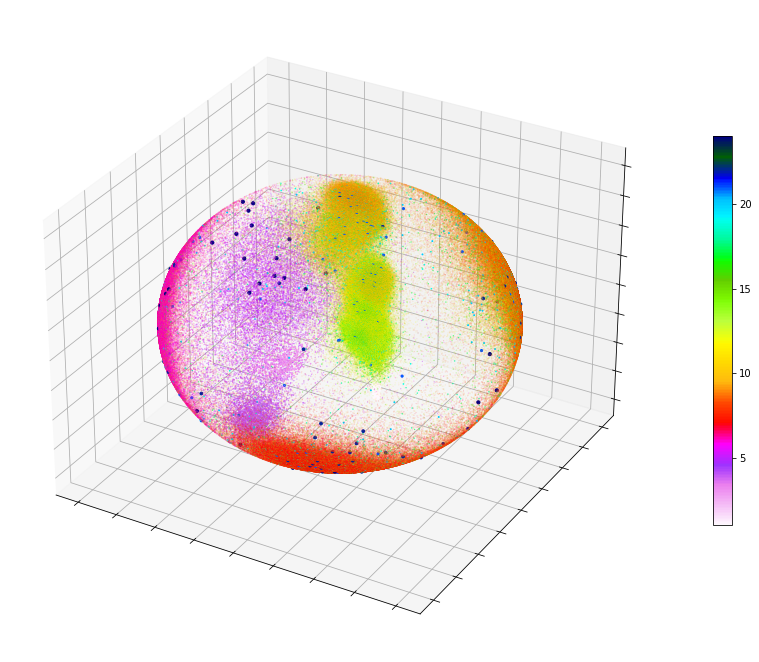}
    \caption{CICY four-folds as visualized by the SNN. 
    The color scheme is $h^{1,1}$.}
    \label{fig:cicy4_cluster}
\end{figure}
The representation of the dataset learned by the Siamese net
in the embedding space $\mathbb{R}^3$ is shown in Figures
\ref{fig:cicy_cluster} and \ref{fig:cicy4_cluster}
where we see that CICY matrices belonging to the same $h^{1,1}$ tend to cluster together.
\subsection{Clustering CICY Manifolds}
We now apply these results to identify subsets of CICYs that are likely to contain
given $h^{1,1}$ values.
Such computations
provide a paradigm to isolate regions of the string landscape by their likelihood to contain 
standard-model-like vacua.
The precise choice of these regions is somewhat subtle, and depends on the confidence with
which we would like to select/reject manifolds in the landscape. 
For illustration, we train a Nearest Neighbors Classifier on the embedding 
space representations learnt by FN for CICY3 and CICY4 respectively. 

Our results in Tables \ref{tab:cicy3exclusion} and \ref{tab:cicy4exclusion} demonstrate that
the SNN hones
into relatively tiny regions of the landscape where these manifolds are most likely to be found. 
In all cases but for $h^{1,1}=21$ for CICY4, the clustering is 
significantly better than random choice.
As an example, the $h^{1,1}=4$ subclass for the CICY4 test set
corresponds to 3829 manifolds
out of the total 901099. Identifying even one manifold from 
this subset correctly by random guessing is extremely unlikely. 
In contrast, the above classifier
predicts 4247 manifolds as corresponding to $h^{1,1}=4$, which is 
correct for 2668 of these. Thus, the learned similarity
score dramatically reduces the search space of $h^{1,1}=4$ CICY4s 
from 901099 to 4247 `most likely' manifolds.
\begin{table}[]
    \centering
    \begin{tabular}{||c| c| c| c|c||} 
 \hline
 $h^{1,1}$& $T,V$ & Total $(Te)$ & True Positive & Predicted (\% $\sum Te$)  \\ 
 \hline\hline
 0& 3,0 & 19 & 11 & 40 (0.52\%) \\ \hline
 1& 3,0 & 2 & 2 & 4 (0.05\%) \\ \hline
 2& 3,0 & 33 & 33 & 47 (0.61\%)\\ \hline
 3& 3,0 & 152 & 85 & 92 (1.20\%)\\ \hline
 4& 9,1 & 415 & 387 & 393 (5.12\%)\\ \hline
 5& 19,2& 835 & 833 & 842 (10.96\%)\\ \hline
 6& 28,3& 1226& 1226& 1235 (16.08\%)\\ \hline
 7& 32,4& 1427& 1425& 1427 (18.58\%)\\ \hline
 8& 29,4& 1295& 1292& 1294 (16.85\%) \\ \hline
 9& 22,3& 1011& 1007& 1008 (13.12\%) \\ \hline
 10& 14,2& 632& 628& 631 (8.22\%) \\ \hline
 11& 8,1& 363& 359& 360 (4.69\%) \\ \hline
 12& 3,1& 157& 151& 151 (1.97\%) \\ \hline
 13& 3,0& 69& 57& 57 (0.74\%) \\ \hline
 14& 3,1& 19 & 15 & 23 (0.30\%)\\ \hline
 15& 3,0& 13 & 12 & 12 (0.16\%) \\ \hline
 16& 1,0&1 & 1 & 27 (0.35\%) \\ \hline
 19& 3,0& 12 & 4 & 34 (0.44\%)\\ \hline
\hline
\end{tabular}
    \caption{Clustering CICY3 manifolds.
    $(T, V, Te)$ denote the training, validation 
    and test sets respectively.
    The clusterer predicts a number of CICY3s for each $h^{1,1}$ (and we indicate the \% of the total $Te$), of which the true positives are recorded in the third column.
    }
    \label{tab:cicy3exclusion}
\end{table}
\begin{table}[]
\centering
\begin{tabular}{||c| c| c| c|c||} 
 \hline
 $h^{1,1}$&$T,V$ & Total $(Te)$ & True Positive & Predicted (\% $\Sigma\, Te$) \\ 
 \hline\hline
  1 &3,0& 4 & 4 & 50 (0.01\%)\\ \hline
  2 &9,1& 93 & 81 & 186 (0.02\%)\\ \hline
  3 &9,1& 756 & 680 & 1091 (0.12\%)\\ \hline
 4 &17,2& 3829 & 2668 & 4247 (0.47\%)\\ \hline
 5 &60,7& 13497 & 8274 & 14656 (1.63\%) \\ \hline
 6 &160,18& 35524& 26421& 35524 (4.45\%) \\ \hline
 7 &322,36& 71372& 54498& 71372 (9.42\%) \\ \hline
 8 &511,57& 113097& 78916& 134858 (14.97\%) \\ \hline
 9 &650,72& 143719& 94946& 182992 (20.31\%) \\ \hline
10 &681,76& 150690& 73226& 159434 (17.69\%) \\ \hline
11 &597,66& 132033 & 49315 & 125712 (13.95\%) \\ \hline
12 &448,50& 99715& 24092& 99175 (10.63\%) \\ \hline
13 &290,32& 64150& 5669& 23922 (2.65\%) \\ \hline
14 &197,16& 37111& 1920& 9572 (1.06\%) \\ \hline
15 &87,9& 19136& 173& 1553 (0.17\%) \\ \hline
16 &42,5& 9505& 168& 7740 (0.86\%) \\ \hline
17 &19,2& 13270& 97& 4214 (1.47\%) \\ \hline
18 &9,1& 1984& 30& 6126 (0.24\%) \\ \hline
19 &9,1& 732& 7& 6161 (0.68\%) \\ \hline
20 &9,1& 343& 5& 6787 (0.75\%) \\ \hline
21 &9,1& 65 & 0 & 2645 (0.29\%)\\ \hline
22 &9,1& 40 & 1 & 2628 (0.29\%)\\ \hline
24 &9,1& 30 & 2 & 1354 (0.15\%)\\ \hline
 \hline
\end{tabular}
    \caption{Clustering CICY4; notation as in Table \ref{tab:cicy3exclusion}.}
    \label{tab:cicy4exclusion}
\end{table}
\subsection{Typical $h^{2,1}$s for CICY3s}\label{sec:typicalcicy3}
The SNN has been trained above on a particular criterion of similarity, namely, the matching of $h^{1,1}$.
We now examine how the clustering learned by the SNN may be interpreted to reflect a still broader notion of similarity in the dataset. To define this question,
we start from the general expectation that manifolds close
to the centroid of a cluster may be thought of as being the `most similar' to all other manifolds in the
cluster. 

\begin{figure}[]
    \centering
    \includegraphics[width=0.4\textwidth]{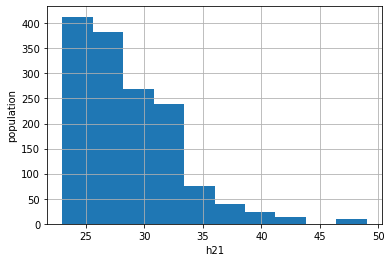}
    \caption{The $h^{2,1}$ populations in the $h^{1,1}=7$ cluster.}
    \label{fig:h21distrib}
\end{figure}

For concreteness, we next examine if $h^{2,1}$ - this is another important Hodge number, but has much wider spread in value - of these manifolds are sufficiently generic to the
$h^{1,1}$ subset or are outliers.
As an example, consider $h^{1,1}=7$ which has 1463 elements.
The 23 distinct $h^{2,1}$ values in this subset
lie in the range $28.45\pm 4.41$ at one confidence level. The complete distribution is visualized in
Figure \ref{fig:h21distrib}.
Analyzing this
clustering using the $k$ means algorithm \cite{lloyd1982least,forgy1965cluster}
picks out $h^{2,1}=26,28,29$, which may indeed be regarded as being `typical'.
We show further details
and corresponding results for other $h^{1,1}$ values in Appendix \ref{app:typicalh21}.
\section{Conclusions and Outlook}
In sum, AI/ML formulations of similarity condense the
vast string landscape into comprehensibility, most notably by providing a 
concrete representation of
the landscape with similar vacua clustered together.
Central to this is \textit{few shot learning}, the ability to draw 
inferences across the full landscape from very limited input data.
This allows us to narrow the search space for `desirable'
vacua by drawing on a very limited training set. This opens up the exciting possibility of
using these methods to obtain strong priors for regions likely to contain standard-model like vacua, thereby
aiding in the search and classification of such solutions.

We further expect this line of exploration to yield still deeper insights into the underlying structures
of the string landscape. Firstly, since \textit{any} compactification scenario 
can be expressed 
as some numerical tensor input as shown here for the CICYs, 
our framework is very generally applicable to the string landscape.
Secondly, identifying criteria for similarity is itself an important advance on classification: 
rather than grouping objects into different categories, we
identify the principal features which allow this grouping to take place. This step is 
the gateway leading from empiricism to understanding. 

Additionally, while we are unaware of a mathematically rigorous framework by which manifolds with different
Hodge numbers may be compared for similarity, our results enable us to do precisely this. Indeed, 
Figures \ref{fig:score_h11} and \ref{fig:score_h11_4fold} indicate that the SNN learns
to regard manifolds with closer values of $h^{1,1}$ as being `more similar' to each other. This was not
an input to the SNN, which is not even shown the actual $h^{1,1}$ values, and must be regarded as 
a nontrivial output. Interestingly, the similarity score is very intuitive for a human
looking at Riemann surfaces
embedded in three dimensions. One would indeed be led to regard surfaces with 3 holes being more similar
to ones with 4 holes than they are to ones with 10 holes. 

Finally, our work also explicitly realizes a general paradigm for
conjecture formulation using AI/ML \cite{He:2021oav}, namely, the ability to generalize significantly beyond
the dataset on which the algorithm is trained. By training on a vanishingly small subset of the landscape 
and extracting meaningful results on the full dataset, we demonstrate that the ability to extract
precise conjectures about the string landscape is well within grasp, even though the full landscape may not
yet be. Hence we expect the analysis here to be only the precursor to an exhaustive exploration of
the string landscape and more generally the physical and mathematical properties of string theory.
\paragraph*{Acknowledgements:} 
YHH would like to thank STFC for grant ST/J00037X/1.
SL acknowledges support from the Simons Collaboration on the Non-perturbative
Bootstrap and the Faculty of Sciences, University of Porto while the work was conceived and partially
carried out.
\section*{Appendices}
\appendix
\section{Calabi Yau Manifolds}\label{ap:histo}
We here give the reader a rapid initiation to Calabi-Yau manifolds; for a recent pedagogical introduction, q.~v.~\cite{he2021calabi}.
The typical student of physics is inculcated to the differential-geometric definition of a manifold $M$, before, if at all, being exposed to the algebro-geometric.
This, in some sense, is reverse in complexity.
One is familiar with (real, affine) algebraic geometry since school Cartesian coordinates.
For instance, the intersection of a linear and a quadratic polynomial in real coordinates $(x,y) \in \IR^2$ prescribes the algebraic variety obtained from the intersection of a line and a conic section, which is generically 2 distinct points in the real plane.

The purpose of (complex) algebraic geometry is to realize manifolds $M$ as polynomials in an ambient space $A$, typically a complex projective space $\IP^n$.  
Because $\IP^n$ is K\"ahler and compact, this ensures that $M$ is also, which is perfect for
string compactification.
In addition, when $M$ has vanishing first Chern class, $M$ will be Calabi-Yau. The point is that such statements as vanishing Chern class, which, in the language of differential geometry, would involve curvature and tensors, but in that of algebraic geometry, involves no more than properties such as degrees of polynomials.

The algebro-geometric set-up gives a completely algebraic (i.e., polynomial and combinatorial) way of constructing a Calabi-Yau manifold, which is precisely why it is perhaps more amenable to machine-learning.
The archetypical example is to take a single polynomial (hypersurface) of homogeneous degree 5 in $\IP^4$ with homogeneous coordinates $[z_0:z_1:z_2:z_3:z_4]$ as
\begin{equation}
   \{ 0 =  \sum\limits_{\alpha}
   C_{\alpha_0,\alpha_1,\alpha_2,\alpha_3,\alpha_4}
   z_0^{\alpha_0}z_1^{\alpha_1}z_2^{\alpha_2}z_3^{\alpha_3}z_4^{\alpha_4}
   \} \subset \IP^4 \ ,
\end{equation}
where $\sum\limits_{i=0}^4 \alpha_i = 5$ and $\alpha_i \in \IZ_{\geq 0}$ so that each monomial is degree 5 and the coefficients $C_{\alpha_0,\alpha_1,\alpha_2,\alpha_3,\alpha_4}$ dictate the compelx structure (shape) of $M$. This is the quintic Calabi-Yau 3-fold.
In general a degree $n+1$ polynomial in $\IP^{n}$ is a compact Calabi-Yau $(n-1)$-fold.
Now, topological quantities do {\it not} depend on $C_{\alpha_0,\alpha_1,\alpha_2,\alpha_3,\alpha_4}$ (and thus do not depend on the detailed monomials, so long as one can choose a generic enough set of monomial terms with generic enough choice of coefficients) so for our purposes the single number 5 suffices to characterize this Calabi-Yau manifold. 

We can generalize by considering $M$, of complex dimension $D$, as intersections of complex-valued polynomials in the homogeneous coordinates of a product $A =  \IP^{n_1} \times \ldots \times \IP^{n_m}$ of complex projective spaces $\IP^{n_i}$. 
When the number of polynomials $K$ is equal to the co-dimension $n_1 + \ldots n_m - D$, so that each new polynomials slices out one complex dimension, we call $M$ a complete intersection Calabi-Yau manifolds (CICY).

\subsection{Complete Intersection Calabi-Yau: CICY}
In brief, a CICY is the matrix for $q_i^j \in \mathbb{Z}_{\geq 0}$
\begin{equation}\label{eq:cicyconf}
\begin{array}{ccc}
    M = 
      \left[\begin{array}{c|cccc}
      n_1 & q_{1}^{1} & q_{1}^{2} & \ldots & q_{1}^{k} \\
      n_2 & q_{2}^{1} & q_{2}^{2} & \ldots & q_{2}^{k} \\
      \vdots & \vdots & \vdots & \ddots & \vdots \\
      n_m & q_{m}^{1} & q_{m}^{2} & \ldots & q_{m}^{k} \\
      \end{array}\right]\,
      &
      &
      \begin{array}{l}
      \sum\limits_{r=1}^m n_r = k+D\\
  	\sum\limits_{j=1}^k q_{i}^{j} = n_i + 1   \\
  	\forall i = 1, \ldots, m \ .
      \end{array}
      \end{array}
\end{equation}
We can thus see \eqref{cicyconf} as the definition of a K\"ahler manifold of complex dimension $D$, as a complete intersection of $k$ polynomials in  $A =  \IP^{n_1} \times \ldots \times \IP^{n_m}$.
Indeed, $q_i^j$ specify the degree of homogeneity of the $j$-th defining polynomial in the homogeneous coordinates of the $i$-th projective ambient space factor.
The complete intersection condition is then $\sum\limits_{r=1}^m n_r = k+D$.
The Calabi-Yau condition (vanishing of the first Chern class) is then
$\sum\limits_{j=1}^k q_{i}^{j} = n_i + 1, \ \forall i = 1, \ldots, m$, so the first column of $n_i$ is redundant information.
Clearly, independent row and column permutations of $M$ define the same manifold.
The coefficients of the defining polynomials are the complex structure parameters and the computation of any topological quantity is independent of these, thus the degree information given by the matrix $M$ suffices.
The quintic example above is then the $1 \times 1$ matrix $[5]$. 

The most important topological quantity of a Calabi-Yau manifold is the list of its Hodge numbers $h^{p,q}$.
The Betti numbers $b^i = \sum\limits_{p+q=i} h^{p,q}$ count the number of $i$-dimension ``holes'' in $M$ and $\sum\limits_{i=0}^D (-1)^i b^i = \chi$ is the Euler number.
For smooth, connected and simply connected Calabi-Yau 3-folds, the only non-trivial Hodge numbers are $h^{1,1}$ and $h^{2,1}$.
For 4-folds, they are $h^{1,1},\,h^{2,1},\,
h^{2,2}$ and $h^{3,2}$, satisfying the constraint $h^{2,2} = 2(22 + 2h^{1,1} + 2h^{3,1} - h^{2,1})$.
In this letter, we are concerned with the complete intersection Calabi-Yau 3-folds and 4-folds, where $D=3,4$, which we denote as CICY3 and CICY4.

\section{CICY Data for the SNN}
We now turn to an overview of the demographics of the CICY datasets along with a discussion of how
the data is prepared for training the SNN. As remarked in the main text, the
population distribution of these datasets is heavily skewed, with densely populated middles and sparsely
populated tails. For example, there is only one CICY3 manifold that corresponds to $h^{1,1}=16$, 5 manifolds
corresponding to $h^{1,1}=1$, while there are 1463 manifolds corresponding to $h^{1,1}=8$. In the same 
vein, there are 7 CICY4 manifolds with $h^{1,1}=1$ and 151447 manifolds with $h^{1,1}=10$. 
\begin{figure}[]
    \centering
    \includegraphics[width=0.4\textwidth]{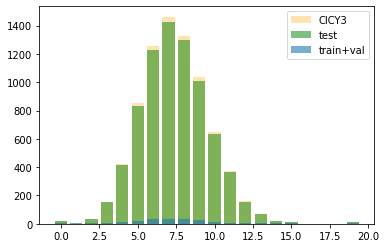}
    \caption{The CICY3 population stratified by $h^{1,1}$ values.}
    \label{fig:pop_h11}
\end{figure}
\begin{figure}[]
    \centering
    \includegraphics[width=0.4\textwidth]{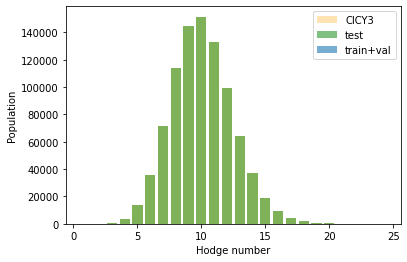}
    \caption{The CICY4 population stratified by $h^{1,1}$ values.}
    \label{fig:pop_h11_4fold}
\end{figure}
This is shown explicitly in Figures \ref{fig:pop_h11} and \ref{fig:pop_h11_4fold}, containing
the histograms of the number of CICY3s and CICY4s for each $h^{1,1}$ along with the train test splits
mentioned below in Appendix \ref{app:traintest}.
We see that the distribution is strongly peaked
around the middle ($h^{1,1} = 5-10$) and dies off sharply at the tails. 
\subsection{Splitting into Train and Test Sets} \label{app:traintest}
A crucial part of the machine learning methodology is to partition the data into training, validation and
test splits. We describe briefly the \textit{raison d'etre} for each of these and our procedure for 
partitioning. 

First, the \textit{train set (T)} is the subset of the data that the SNN is given to learn from. By using
this data, the SNN tunes the weights of the features network to make optimal decisions about similarity of
a given pair. Next, the \textit{validation set (V)} is the data which is used to periodically evaluate the
SNN while it trains, but is not given to the network to train on. Ideally, the performance of the network
on the train and validation sets should be comparable. Finally, the \textit{test set (Te)} is the complement
of $T$ and $V$ in the dataset. This is not shown to the SNN until after training is completed as is used
to evaluate the performance of the network, and explicate the extent to which the SNN
has solved the given problem. A typical choice for the partitioning of the data 
could be $(0.6,0.2,0.2)$, i.e., 60\% of the data is used for training, 20\% for validation, 
and 20\% for testing, and often this partitioning is done by 
\textit{random sampling}, i.e. we pick random subsets of the dataset in these fixed proportions.

However, completely random sampling in imbalanced datasets is not always possible; one may end up with
zero elements of sparsely populated classes in one or more of the $T,V,Te$ subsets. We therefore 
carry out \textit{stratified} random sampling, i.e. we partition each $h^{1,1}$ class in the CICY datasets
into $T,V,Te$ subsets by random sampling
and concatenate these to arrive at the full training, validation, testing data. This also enables us to
drive down the size of the $T+V$ subset while ensuring elements are drawn from each $h^{1,1}$ class.
In this work we have split the CICY3 dataset as $(0.0225,0.0025,0.975)$, subject to a minimum
of 3 elements in $T+V$. 
As an illustration, consider the $h^{1,1}=4$ class in CICY3, which has 425 elements. The above split
yields 9 elements in $T$, 1 in $V$ and 415 in $Te$. 
The CICY4 dataset on the other hand is split as $(0.0045,0.0005,0.995)$, subject to a minimum of 10 elements
in $T+V$. Consider as an example the $h^{1,1}=2$ class which has 103 elements. Splitting the data as above,
without a minimum would lead to zero elements in $T+V$ from this class, hence the minimum cap.

The complete
partitioning for the CICY3 dataset is shown in the second and third columns of 
Table \ref{tab:cicy3exclusion} and for CICY4 in the corresponding columns of Table \ref{tab:cicy4exclusion},
as well as in 
the histograms of Figures \ref{fig:pop_h11} and \ref{fig:pop_h11_4fold}.

\subsection{Feature Engineering}\label{app:feature_engg}
\begin{figure}[]
  \centering
  \subcaptionbox{Original $3\times 4$
  \label{cicyorig}}{\includegraphics[height=1.4in]{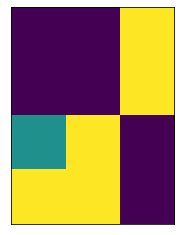}}\hspace{1em}%
  \subcaptionbox{padded to $18\times 18$
  \label{cicyups}}{\includegraphics[height=1.4in]{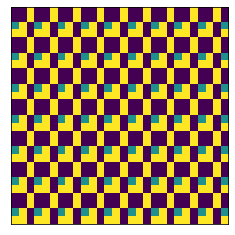}}
  \caption{CICY4 configuration matrices as images.}
\label{fig:cicy4images}
\end{figure}

The CICY data is in the form of matrices of variable shape, with integer entries 
ranging from 0 to 5 in CICY3 and 0 to 6 in CICY4. To make the data more amenable to machine learning we
firstly resize the matrices to a uniform $n\times n$ by padding. In practice, we find good results by
padding the CICY3 matrices to size $18\times 18$ by appending constant values $-1$ on each side until the
desired size is reached. A toy example of a $2\times 2$ matrix padded to $4\times 4$
in this manner is
\begin{equation}
 \begin{pmatrix} a& b\\c& d\end{pmatrix}
\Rightarrow 
\begin{pmatrix}     -1 & -1& -1& -1\\
-1 & a& b& -1\\    -1 & c& d& -1\\    -1 & -1& -1 &-1\\\end{pmatrix}\,.
\end{equation}
As an explicit example, we have shown the image representation of a $13\times 15$ matrix from CICY3
as well as the corresponding padded $18\times 18$ matrix in Figures \ref{cicyorig}
and \ref{cicyups} respectively.
\begin{figure}[]
  \centering
  \subcaptionbox{Original $13\times 15$
  \label{cicyorig}}{\includegraphics[height=1.4in]{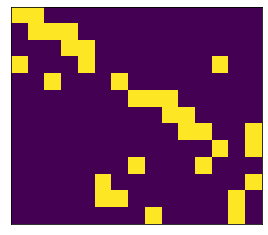}}\hspace{1em}%
  \subcaptionbox{padded to $18\times 18$
  \label{cicyups}}{\includegraphics[height=1.4in]{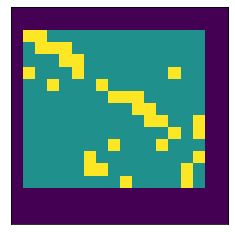}}
  \caption{CICY3 configuration matrices as images.}
\label{fig:cicy3images}
\end{figure}
In contrast, for the CICY4 matrices better results are obtained by wrapping the original matrix entries
in all four directions until the desired size is reached. This yields, for the same $2\times 2$ 
example,
\begin{equation}
 \begin{pmatrix} a& b\\c& d\end{pmatrix}
\Rightarrow 
\begin{pmatrix}     d & c& d& c\\
b & a& b& a\\    d & c& d& c\\    b & a& b &a\\\end{pmatrix}\,.
\end{equation}
A CICY4 matrix before and after padding in this manner is shown in Figure \ref{fig:cicy4images}.
This uniformization of the configuration matrix also prevents the neural network from
learning spurious correlations in the dataset. Namely,
a significant fraction $\left(\sim 50\%\right)$
of both the CICY3 and CICY4 datasets are \textit{favourable}, in that
the Hodge number
$h^{1,1}$ equals the number of rows of the configuration matrix. Geometrically this 
means that ambience projective space K\"ahler classes descent completely to the the CICY.
Since the matrices are now uniformized, this correlation is removed from the CICY data. Next, we
rescale the matrix entries \textit{via}
\begin{equation}
    x_{ij} \mapsto \hat{x}_{ij} = 2\times \frac{x_{ij}}{\max(\lbrace x\rbrace)}-1\,,
\end{equation}
where $x_{ij}$ is the $i,j$th entry in a matrix $x$ belonging to a CICY dataset and $\max(\lbrace x\rbrace)$
is the maximum value among all matrix entries in that dataset.
Of course, this scaling does not mean anything in the algebraic geometry because the matrix entries are the multi-degrees of the defining polynomials. However, the scaling is an equivalent representation and the normalized data is more amenable to ML.
\section{Features Network}\label{app:convnet}
\begin{figure}[]
  \centering
  \subcaptionbox{Image}{\includegraphics[width=1.1in]{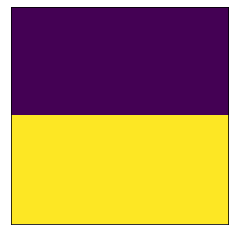}}\hspace{1em}%
  \subcaptionbox{Extracted edge}{\includegraphics[width=1.2in]{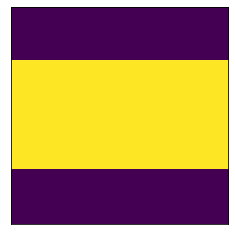}}
  \caption{Extracting horizontal edges from images.}
\label{fig:edge_extraction}
\end{figure}

As mentioned in Section III, the design of the features network FN relies crucially on the incorporation
of Convolutional Layers. These are responsible for efficiently extracting
local patterns in the image, which are then processed further by the neural network. 
For example, a face recognition algorithm would typically extract and compare
patters associated with common landmarks on the face such as eyes, noses, lips etc. This is usually 
accomplished by the means of \textit{filters}, which are matrices that are scanned across the image to
extract particular features, where the entries of the matrices are determined by the kind of feature one
aims to extract. As a simple example, consider a $5\times 5$ image with a horizontal edge across the centre.
This may be represented by the matrix
\begin{equation}
\mathrm{Img}=\begin{pmatrix}     0 & 0& 0& 0& 0&0\\
0 & 0& 0& 0& 0&0\\    0 & 0& 0& 0& 0&0\\    1 & 1& 1& 1& 1&1\\
    1 & 1& 1& 1& 1&1\\     1 & 1& 1& 1& 1&1\\     \end{pmatrix}\,.
\end{equation}
The $3\times 3$ filter
\begin{equation}
    \mathrm{filter} = \begin{pmatrix}-1 & -1 &-1\\ 0 & 0 &0\\
    1& 1 &1 \end{pmatrix}
\end{equation}
is used to extract the horizontal edge from the above image by means of convolutions.
The convolution operation involves sliding the filter over the image 
(in steps of 1 for simplicity), computing $*$ -- the element-wise/Hadamard product --
and summing all elements of the 
resulting matrix. As an example, consider the $3\times 3$ submatrix of the image with the
$1,0$ element on the upper left corner. The above operation yields
\begin{equation}
    \Sigma\left[\begin{pmatrix}0 & 0 &0 \\
    0 & 0 &0\\ 1& 1 &1 \end{pmatrix}* 
    \begin{pmatrix}-1 & -1 &-1\\ 0 & 0 &0\\
    1& 1 &1 \end{pmatrix}\right] 
    = \Sigma \begin{pmatrix}0 & 0 &0\\ 0 & 0 &0\\ 1& 1 &1 \end{pmatrix}=3\,.
\end{equation}
Applying this convolution to the whole matrix yields
\begin{equation}
\mathrm{Img}'=\begin{pmatrix}    0 & 0& 0& 0\\    3 & 3& 3& 3\\    3 & 3& 3& 3\\
     0 & 0& 0& 0\\    \end{pmatrix}\,,
\end{equation}
i.e. the horizontal edge has been extracted successfully. This is clearly visible from the images 
corresponding to $\mathrm{Img}$ and $\mathrm{Img}'$ shown in Figure \ref{fig:edge_extraction}.
Extracting a single feature is typically insufficient to characterize an image; multiple features
are needed. This requires passing the image through multiple filters.
In the above example, since the form of the feature
was simple, an appropriate filter could easily be constructed.
In general, the identification of appropriate filters 
corresponding to complicated features is a notoriously difficult
problem. Indeed, even identifying the optimal set of features to characterize images on
for a dataset is a far from obvious task.

A central insight of deep learning to computer vision is that 
rather than using predetermined filters, 
we should instead treat the matrix entries of filters as 
tunable parameters to be optimized on the given dataset \cite{lecun1989backpropagation}. This is 
accomplished by incorporating a convolutional layer in the neural network, which is essentially a stack
of tunable filters. This allows the neural net to determine in one go both the optimal features to 
classify along and the appropriate filters for doing this classification.
A neural network built from convolutional layers is called a convolutional neural network
or a ConvNet.

\section{Training the SNN}\label{s:training}
\begin{figure}[h]
    \centering
    \includegraphics[width=0.45\textwidth]{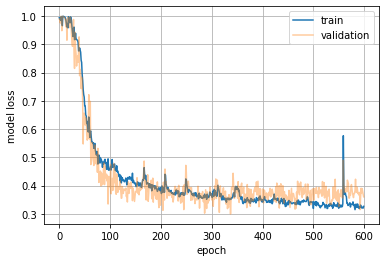}
    \caption{The loss curve for the Siamese Net trained on similarity with respect to
    $h^{1,1}$ values of CICY3 matrices.}
    \label{fig:training_h11}
\end{figure}

\begin{figure}[h]
    \centering
    \includegraphics[width=0.45\textwidth]{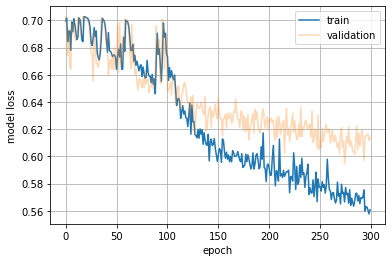}
    \caption{The loss curve for the Siamese Net trained on similarity with respect to
    $h^{1,1}$ values of four-folds.}
    \label{fig:training_h11_4fold}
\end{figure}

The SNN is trained with the triplet loss function and the \texttt{Adam} optimizer with a learning rate of
0.01. The resulting loss curves for the train and validation sets are shown in 
Figures \ref{fig:training_h11} and \ref{fig:training_h11_4fold}. The performance on both sets is comparable
throughout training.
The model with the minimum validation loss is then chosen for evaluation.

\section{Typical $h^{2,1}$s for CICY3s}\label{app:typicalh21}
As a final addendum, we provide details on how the $h^{1,1}$ clusters in CICY3 are analyzed to compute 
`typical' values of $h^{2,1}$ corresponding to them, as outlined in 
Section \ref{sec:typicalcicy3}. For definiteness, we focus on the $h^{1,1}=7$ cluster for which the
final result has already been mentioned. 
The analysis is exactly analogous for the remaining clusters.
We analyze these clusters using $k$-means clustering, an unsupervised learning algorithm, reviewed here.

A $k$-means clusterer organizes a point cloud into $k$ clusters by determining the 
location of the centroid of each cluster and assigning each point in the point cloud to the cluster with
the centroid closest to it. Here $k$ is an external parameter which must be supplied to 
the algorithm.

\begin{figure}[]
    \centering
    \includegraphics[height=0.35\textwidth, width =0.4\textwidth]{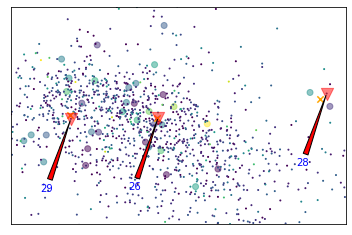}
    \caption{The $h^{1,1}=7$ cluster, with the $h^{2,1} = 26, 28, 29$ marked. }
    \label{fig:h21cluster}
\end{figure}

\begin{figure}[]
    \centering
    \includegraphics[height=0.30\textwidth,width=0.45\textwidth]{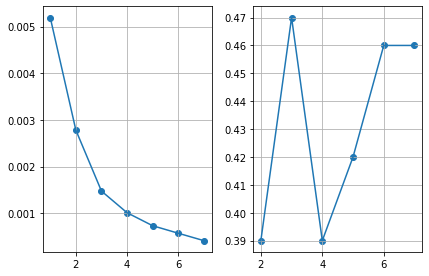}
    \caption{The inertia (left) and silhouette score (right) for the $h^{1,1}=7$
    CICY3 
    cluster. 
    The $x$ and $y$ axes are the $k$ values and corresponding scores respectively.}
    \label{fig:kmeanscicy3}
\end{figure}

\begin{table}[]
    \centering
    \begin{tabular}{||c| c| c| c||} 
 \hline
 $h^{1,1}$& median $h^{2,1}$ & $\bar{h^{2,1}}\pm\sigma_{h^{2,1}}$ & Typical $h^{2,1}$  \\ 
 \hline\hline
 4& 42 & $41.07\pm 7.65$ & 41, 43\\ \hline
 5& 35 & $36.12\pm 6.22$ & 33, 33, 41, 48\\ \hline
 6& 31 & $31.75\pm 5.29$& 30, 30\\ \hline
 7& 28 & $28.45\pm 4.41$& 26, 28, 29\\ \hline
 8& 25 & $25.99\pm 3.76$& 28, 30 \\ \hline
 9& 23 & $23.73\pm 3.00$& 27, 30\\ \hline
 10& 21& $21.95\pm 2.56$& 20, 20\\ \hline
 11& 19& $20.13\pm 2.10$& 19, 19, 21\\ \hline
\hline
\end{tabular}
    \caption{Typical $h^{2,1}$ values for given $h^{1,1}$ clusters.}
    \label{tab:typicalh21cicy3}
\end{table}

A convenient rule of thumb for selecting the optimal value of $k$ is the elbow rule.
Begin by defining the \textit{inertia} of the clustering, the mean square distance of the point cloud
elements to the centroid closest to them. Clearly, a good clustering would be associated with a low value
of inertia. Increasing $k$ allows us to add more centroids, in turn increasing
the likelihood for every point to have a centroid close to it. This tends to drive down the inertia.
Typically, however, the inertia falls rapidly until an optimal value of $k$ is reached, after which its
rate of decrease is much less. That is, there is little apparent benefit to adding more centroids.
This is visualized as an `elbow' in the inertia vs $k$ graph, and the
optimal value of $k$ is the location of the elbow.

This graph is plotted on the left
in Figure \ref{fig:kmeanscicy3}
for the $h^{1,1}=7$ and we see clearly that the elbow is located at $k=3$. 
This optimal value may be further cross-checked by computing the silhouette score, which is the mean
silhouette coefficient, defined by 
\begin{equation}
    \frac{b-a}{\max(a,b)}\,,
\end{equation}
across the dataset. Here $a$ is the mean intra-cluster distance of the given point, while $b$ is the 
mean distance from points in the nearest cluster. When the point is near the centre of the cluster, 
$b>>a$ and the coefficient is nearly 1. In contrast, when the point is near the 
edge, 
$b\simeq a$ and the
score is nearly 0. For a point assigned to the wrong cluster, $b<<a$ and the score is nearly $-1$. 
These observations suggest that the 
silhouette score should then ideally reach a maximum for the optimal value of $k$. This analysis again
yields $k=3$.
It is then straightforward to fit a $k$-means 
clusterer with $k=3$ on the train set, fit it to the test set,
identify the manifold closest to the centroid of each point cloud and read
off the corresponding $h^{2,1}$ values. This yields 26, 28 and 29 as mentioned in Section \ref{sec:typicalcicy3}.
The cluster is shown in Figure \ref{fig:h21cluster} after projection to two dimensions using
principal
component analysis. The orange crosses indicate 
the centroids and the red triangles the typical CICY3 manifolds detected. The tails of the red
arrows
pointed at the typical CICYs carry the corresponding $h^{2,1}$ values.
The larger points denote the 
manifolds in the training set while the smaller points denote the manifolds in the test set.
The colours of the points correspond to their $h^{2,1}$ values.

We may repeat this analysis with $h^{1,1}$ ranging from 4 to 11. There are more than 350 manifolds for
each case, a number large enough that the notion of `typical' is meaningful. 
Our results are summarized in Table \ref{tab:typicalh21cicy3}.
Note that the number of typical CICYs is different for different $h^{1,1}$ values. This is due to the
variance in the number of clusters detected for each $h^{1,1}$ point cloud.
\bibliographystyle{apsrev4-1}
\bibliography{references}
\end{document}